# MGNN: Moment Graph Neural Network for Universal Molecular Potentials

Jian Chang[1], Shuze Zhu*[1]

[1]Center for X-Mechanics, Institute of Applied Mechanics,

Zhejiang University, Hangzhou 310000, China

*To whom correspondence should be addressed. E-mail: shuzezhu@zju.edu.cn

**Abstract**

The quest for efficient and robust deep learning models for molecular systems representation is increasingly critical in scientific exploration. The advent of message passing neural networks has marked a transformative era in graph-based learning, particularly in the realm of predicting chemical properties and expediting molecular dynamics studies. We present the Moment Graph Neural Network (MGNN), a rotation-invariant message passing neural network architecture that capitalizes on the moment representation learning of 3D molecular graphs, is adept at capturing the nuanced spatial relationships inherent in three-dimensional molecular structures. MGNN demonstrates new state-of-the-art performance over contemporary methods on benchmark datasets such as QM9 and the revised MD17. The prowess of MGNN also extends to dynamic simulations, accurately predicting the structural and kinetic properties of complex systems such as amorphous electrolytes, with results that closely align with those from ab-initio simulations. The application of MGNN to the simulation of molecular spectra exemplifies its potential to significantly enhance the computational workflow, offering a promising alternative to traditional electronic structure methods.

# Introduction

The computational determination of molecular properties has been significantly accelerated by machine learning. The crux of this advancement lies in the development of machine learning potentials with remarkable efficiency and fidelity. Although the usage of local descriptors in most machine learning interatomic potentials have improved parallel computation and scalability,[1–6] yet they fall short in accuracy compared with atom-centered message-passing neural network (MPNNs) interaction potentials within the concept of graph neural network (GNN),[7–12] where the molecular systems are treated as 3D graphs, with atoms as nodes linked by edges within a defined cutoff radius. These networks have achieved groundbreaking results, often surpassing traditional models with minimal manual feature engineering. They encode atoms into a high-dimensional space and model interatomic interactions through message passing, effectively serving as surrogates for quantum mechanical simulations.[7–16]

In principle, to accurately predict scalar quantities like potential energy, vectorial properties such as forces and dipole moments, and even tensorial properties like polarizabilities[17–20], a GNN which fully accommodates the transformation properties of scalar or tensor is required. For instance, while the translation and rotation of a small molecule do not alter its formation energy, the forces acting on its atoms rotate with the molecule. Most available GNN machine learning potentials address above points by empirically constructing neural network using geometrical quantities like interatomic-distances, bond angles.[21–24] Although they can address the transformation properties to certain extent, they are not based on a rigorous mathematical foundation to ensure that all possible molecular properties satisfying all the needed symmetries can be learned, which might impose a limitation on their predictive power. A systematical, rigorous and universal routine to treat symmetries is highly desirable.

To address the above challenges, in this work we introduce the Moment Graph Neural Network (MGNN), which innovatively propagates information between molecular systems using moments—quantities that encapsulate the spatial relationships between atoms. Inspired by Moment Tensor Potential[4], which has rigorous mathematical proof that any regular function satisfying all the needed symmetries can be approximated while being computationally efficient, our framework defines molecular moments[4] to convey relative spatial relationships between atoms within a molecule, utilizing Chebyshev polynomials to encode interatomic distance information instead of more commonly used functions like Bessel and Gaussian radial basis functions.[21,25] Impressively, our MGNN architecture has achieved new state-of-the-art performance on popular benchmarks like the QM9 dataset[26,27] and the revised MD17 dataset[28]. Additionally, we demonstrate the high consistency of our framework in predicting molecular dynamics simulations of amorphous electrolytes. Moreover, MGNN's predictive capabilities extend to the calculation of dipole moments and polarizabilities for organic molecules, such as ethanol, enabling the rapid simulation of molecular spectra in vacuum conditions.

# MGNN Architecture

## 1.1. Constructing MGNN

### 1.1.1. Overview

The Hamiltonian of a molecular system is uniquely determined by the external potential, depending on the charge quantities $\{Z_i\}$ and atomic positions $\{\vec{r}_i\}$. Formally, a molecule in a specific conformation can be represented as a point cloud with $n$ atoms denoted as $M = (Z, R)$, where $Z = (Z_1, \ldots, Z_n) \in \mathbb{Z}^n$ represents the atom type vector and $R = (\vec{r}_1, \ldots, \vec{r}_n) \in \mathbb{R}^{n \times 3}$ represents the atom coordinate matrix.

MGNNs treat molecules as graphs, with atoms as nodes, and edges representing connections between atoms, either based on predefined molecular graphs or atoms connected within a cutoff distance $r_{cut}$. MGNNs represent each atom through atomic embeddings $h_i \in \mathbb{R}^{H_a}$ where $H_a$ is the number of atom features and the edges between nodes $i$ and $j$ through edge embeddings $e_{ij} \in \mathbb{R}^{H_e}$ where $H_e$ is the number of edge features. These embeddings are updated in each layer through information propagation. In general, the Message Passing Neural Networks (MPNN)[14] framework achieves message passing by $m_v^{t+1} = \sum_{w \in \mathcal{N}(v)} M_t(h_v^t, h_w^t, e_{vw})$ and updates by $h_v^{t+1} = U_t(h_v^t, m_v^{t+1})$, where $M_t$ represents the message function, $U_t$ represents the vertex update function, and $\mathcal{N}(v) = \{w | \|\vec{r}_{vw}\| \leq r_{cut}\}$ denotes the neighbors of node $v$ in graph $G$. Our MGNN framework utilizes moment interaction to construct $M_t$ and $U_t$.

In this work, a molecular graph is represented as $G = (A, E, T)$, where $A$ denotes the atoms (nodes), $E$ the edges within cutoff $r_{cut}$, and $T$ the triplets. Each atom $i$ forms an edge with every neighboring atom within its cutoff distance, as demonstrated in **Figure 1**A, constructing edges $ij$, $ik$, $im$, and corresponding triplets $ijk$, $ijm$, $ikm$ to formulate the moments. Within a triplet composed of nodes $i, j, k$ and edges $ij$, $ik$, the message of triplets are used to update edge representation features, which are further used to update node features. As shown in **Figure 1**B, for a triplet $ijk$ whose center is atom $i$ and edges are $ij$ and $ik$, the information within this triplet first passes to edge $ij$ (symmetrically to $ik$), and then the message of edge $ij$ passes to the central node $i$.

The MGNN framework comprises an embedding block, several sequential interaction layers, and an output block, as shown in **Figure 1**C. The molecular charge quantities (labeled as $Z$) first pass through the embedding block to obtain the initial node representation vectors (labeled as $X$). These initial vectors, along with the positions of atoms in Cartesian space containing geometry information (labeled as $R$), are then input into an interaction layer, which uses an interaction block that generates a correction ($\Delta X$) to be added to the input ($X$) to generate the current layer output (i.e., $X + \Delta X$), which serves as the input for the next interaction layer or output layer. The interaction block (detailed in **Figure 1**D) captures how atoms interact with each other, by processing the message from the neighbors of atom $i$, during which the interaction between edges from atom $i$ (i.e., triplet moment interaction as detailed in **Figure 1**E) is involved. There are $N$ sequential interaction layers in total. The output after the last interaction layer is labeled as $X^L$, which is then fed into target-dependent output block (detailed in **Figure 2**). By aggregating the message of triplets, we construct a comprehensive representation of the molecular structure within the graph, which is then propagated through the interaction layers to update atomic representations. The resulting feature vectors for each atom are fed into the output block, tailored to predict specific properties, culminating in the model's final predictions. The output module of the MGNN model is adept at generating scalar, vector, and tensor outputs, demonstrating its versatility in capturing and predicting a wide range of molecular properties.

The definition of MLP in current work (as shown in **Figure 1** and **Figure 2**) is $MLP(x) = W^{out}\left(\sigma(W^{in}x + b^{in})\right) + b^{out}$, where $W^{in}$ and $W^{out}$ are learnable weights and $b^{in}$ and $b^{out}$ are learnable biases, and $\sigma(x)$ denotes the element-wise nonlinear activation function. In our experiments, we utilize the SiLU[29,30] and Mish[31] activation functions. The action of "*Concat*", "*Reshape*", "*Split*" in current work (as shown in **Figure 1** and **Figure 2**) are used to manipulate data dimensions. All the notations "$W$" and "$b$" with or without superscript are learnable parameters and we will hint their shape when necessary. Next, we will provide a detailed introduction to each block. The notions employed in this paper are summarized in Supporting Information S1.

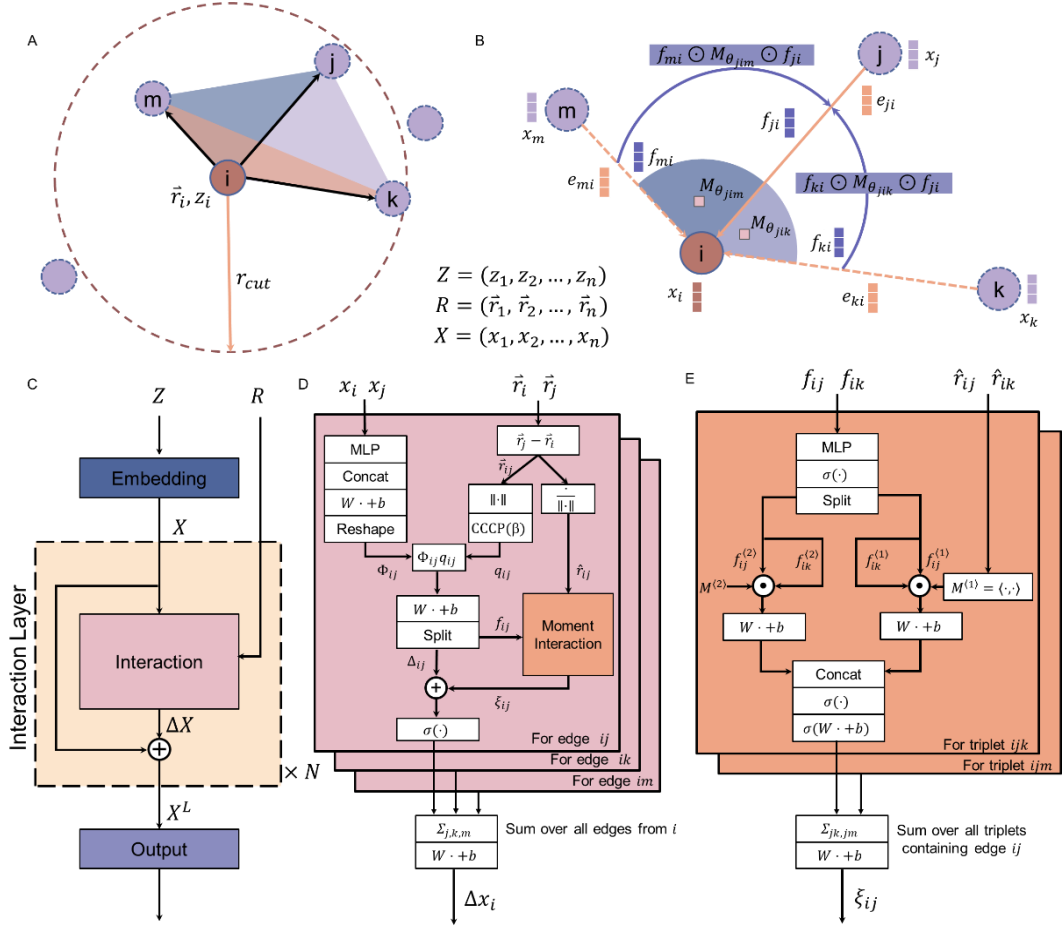

**Figure 1.** The moment message passing scheme (A and B) and the architecture of MGNN (C, D, and E). (A) Atom $i$ receives information from neighboring atoms $j$, $k$, and $m$ within a specified cutoff radius $r_{cut}$, forming three distinct triplets ($ijk$, $ijm$, $ikm$) with $i$ as the central atom. (B) Information flow within a specific triplet $ijk$, where $i$ serves as the central atom and $ij$ and $jk$ constitute the connecting edges. (C) The sequential arrangement of MGNN, from the initial embedding module, through multiple ($N$) interaction layers to the output module. (D) Architecture of interaction layer, which contains moment interaction layer. (E) Architecture of moment interaction layer. "*Concat*" refers to the concatenation of two feature vectors, "*CCCP*" the continuous cutoff and Chebyshev polynomials-based radial basis function, "*Split*" the splitting of a long vector into shorter vectors, "$\oplus$" the residual connection, "$\odot$" the element-wise product, and the dot "·" in the brackets or follows the "$W$" denotes the input from previous block. See main text for detailed description.

*1.1.2. Embedding*

MGNN works on embedded feature, which is initialized by the embedding block shown in **Figure 1**C. Throughout interaction layers, we represent the atoms using a tuple of features $X^l = (x_1^l, \ldots, x_n^l)$, with $x_i^l \in \mathbb{R}^F$ with the number of feature maps $F$, the number of atoms $n$ and the current interaction layer $l$. We maintain a constant number of feature maps at $F = 512$ throughout the network, if not stated otherwise. The representation of atom $i$ is initialized using an embedding depend on the atom type $Z_i \in \mathbb{N}$:

$$x_i^0 = Embedding(Z_i) \tag{1}$$

The atom type embeddings are learned during training.

*1.1.3. Interaction layer*

Here we explain the interaction layer shown in **Figure 1**C by implementing **Equation (2)-(13)**. Within the interaction layer, an interaction block (as depicted in **Figure 1**D) produces a correction $\Delta X = \{\Delta x_i\}$ that is used to update the input embedded feature $X = \{x_i\}$. Formally, the update uses a residual connection inspired by ResNet[32], where $l$ is the layer index, and $\mathcal{N}(i)$ is the neighbor list of atom $i$ that determines which edges and angles are involved in message passing.

$$x_i^{l+1} = x_i^l + \Delta x_i^l\left(x_i^l, \mathcal{N}(i)\right) \tag{2}$$

The residual $\Delta x_i^l$ is computed through integrating information from edges and angles associated with $\mathcal{N}(i)$. For example, for the $\mathcal{N}(i)$ shown in **Figure 1**A, one angular message is obtained from two edge vectors $\vec{r}_{ij}$ and $\vec{r}_{ik}$ originating from node $i$ to node $j$ and $k$ within one triplet $ijk$, another two angular messages are from triplets $ijm$ and $imk$. Since each interaction layer has same structure, for clarity, we no longer indicate the layer index $l$ unless necessary.

In one interaction block (as depicted in **Figure 1**D), the atomistic representations obtained from the preceding layer (embedding block or previous interaction layer) are passed through a learnable MLP layer:

$$h_i = MLP(x_i) \tag{3}$$

After updating the atomistic representations using **Equation (3)**, we concatenate the feature maps of the source and destination nodes of each edge (labeled as $h_i||h_j$ where the symbol "$||$" denotes the concatenation of two vectors) to obtain vectors of dimension $2F$. Subsequently, we pass them through a linear layer to convert the feature size of each edge to dimension $\beta F$, and then reshape them into a matrix $\Phi_{ij} \in \mathbb{R}^{F \times \beta}$ (**Equation (4)**) and next perform matrix multiplication (**Equation (6)**) with a vector $q_{ij} \in \mathbb{R}^{\beta}$ (**Equation (5)**) which encodes the information of $r_{ij}$ using Chebyshev polynomials. The outcome is an edge feature vector $e_{ij} \in \mathbb{R}^F$ (**Equation (6)**):

$$\Phi_{ij} = Reshape(W(h_i||h_j) + b) \in \mathbb{R}^{F \times \beta} \tag{4}$$

$$q_{ij} = \psi^{(\beta)}(r_{ij}) \in \mathbb{R}^{\beta} \tag{5}$$

$$e_{ij} = \Phi_{ij} q_{ij} \in \mathbb{R}^F \tag{6}$$

where $W \in \mathbb{R}^{\beta F \times 2F}$, $b \in \mathbb{R}^{\beta F}$.

The $\psi^{(\beta)}$ in **Equation (5)** is the continuous cutoff and Chebyshev polynomials-based radial basis function ("$CCCP(\beta)$" in **Figure 1**D), denoted as

$$\psi^{(\beta)}(r_{ij}) = \begin{cases} C^{(\beta)}(r_{ij})(r_{ij} - r_{cut})^2, & r_{ij} < r_{cut} \\ 0, & r_{ij} \geq r_{cut} \end{cases} \tag{7}$$

Here $C^{(\beta)}$ are Chebyshev polynomials on the interval $[r_{min}, r_{cut}]$, where $r_{min}$ is the minimal distance between atoms and is usually set to zero, $r_{cut}$ is the cutoff radius which is introduced to ensure a smooth behavior of MGNN when atoms leave or enter the interaction neighborhood. An illustration of the radial basis functions is given in **Figure S1** (Supporting Information S2).

Once the feature maps $e_{ij}$ on the edge are obtained, we split its information (**Equation (8)**) to prepare the update on the features of atoms and edges.

$$\Delta_{ij}, f_{ij} = Split(We_{ij} + b) \tag{8}$$

where $W \in \mathbb{R}^{3F \times F}$, $b \in \mathbb{R}^{3F}$, $\Delta_{ij} \in \mathbb{R}^F$, and $f_{ij} \in \mathbb{R}^{2F}$. The split information is used to compute the incremental node features as in **Equation (9)**.

$$\Delta x_i = W\left(\sum_{j \in \mathcal{N}(i)} \sigma\left(\Delta_{ij} + \xi_{ij}\left(f_{ij}, \vec{r}_{ij}, \mathcal{N}(ij)\right)\right)\right) + b \tag{9}$$

where $W \in \mathbb{R}^{F \times F}$, $b \in \mathbb{R}^F$, and $\xi_{ij}$ represents the moment interaction function, as in **Equation (10)**.

$$\xi_{ij} = W^{out}\left(\sum_{ik \in \mathcal{N}(ij)} \sigma\left(W^{in}\left(m_{jik}^{\langle 1 \rangle} \| m_{jik}^{\langle 2 \rangle}\right) + b^{in}\right)\right) + b^{out} \tag{10}$$

where $W^{in} \in \mathbb{R}^{F \times 2F}$, $b^{in} \in \mathbb{R}^F$, $W^{out} \in \mathbb{R}^{F \times F}$, $b^{out} \in \mathbb{R}^F$, the symbol "$\|$" denotes the concatenation of two vectors. $\mathcal{N}(ij)$ represents the set of neighboring edges of the edge $ij$ that share $i$ as a common vertex for all triplets, as shown in the example in **Figure 1**B, where $\mathcal{N}(ij) = \{ik, im\}$. Note that applying the non-linearity activation function before aggregation improve performance.[33]

In **Equation (10)**, the calculation of $m_{jik}^{\langle v \rangle}$ (i.e., the message passed by triplet $jik$) involves the moments that are the core part of our algorithm, as shown in **Figure 1**B, E, which are calculated as

$$m_{jik}^{\langle v \rangle} = \sigma\left(W\left(f_{ij}^{\langle v \rangle} \odot f_{ik}^{\langle v \rangle} \odot M_{\theta_{jik}}^{\langle v \rangle}\right) + b\right), v = 1,2 \tag{11}$$

where $W \in \mathbb{R}^{F \times F}$, $b \in \mathbb{R}^F$, and "$\odot$" denotes element-wise product.

The $M_{\theta_{jik}}^{\langle v \rangle}$ in **Equation (11)** is the rank $v$ moment of the triplet $jik$, and is calculated as

$$M_{\theta_{jik}}^{\langle v \rangle} = \langle \hat{r}_{ij}^{\otimes v}, \hat{r}_{ik}^{\otimes v} \rangle, v = 1,2 \tag{12}$$

where $\hat{r}_{ij} = \frac{\vec{r}_{ij}}{r_{ij}}$ is the unit vector of $\vec{r}_{ij}$, $\hat{r}_{ij}^{\otimes v} = \underbrace{\hat{r}_{ij} \otimes \cdots \otimes \hat{r}_{ij}}_{v \text{ times}}$ is the Kronecker product of $v$ copies of the vector $\hat{r}_{ij} \in \mathbb{R}^3$ and is a tensor of rank $v$. Notation "$\langle \cdot, \cdot \rangle$" signifies the contraction (product) of tensors, with $v = 1$ corresponding to the vector dot product and $v = 2$ corresponding to the Frobenius product of matrices. In other words, $M_{\theta_{jik}}^{\langle 1 \rangle} = \langle \hat{r}_{ij}, \hat{r}_{ik} \rangle = \|\hat{r}_{ij}\| \times \|\hat{r}_{ik}\| \times \cos(\angle \hat{r}_{ij}, \hat{r}_{ik}) = \cos(\angle \hat{r}_{ij}, \hat{r}_{ik})$, and $M_{\theta_{jik}}^{\langle 2 \rangle} = \langle \hat{r}_{ij}^{\otimes 2}, \hat{r}_{ik}^{\otimes 2} \rangle = \hat{r}_{ij}^{\otimes 2} : \hat{r}_{ik}^{\otimes 2} = \cos^2(\angle \hat{r}_{ij}, \hat{r}_{ik})$, reflecting the angular relationship within the triplet $jik$.

The $f_{ij}^{\langle \nu \rangle}$ in **Equation (11)** is calculated as

$$f_{ij}^{\langle 1 \rangle}, f_{ij}^{\langle 2 \rangle} = Split\left(\sigma\left(MLP(f_{ij})\right)\right) \tag{13}$$

where $f_{ij}$ is calculated using **Equation (8)**, and $f_{ij}^{\langle \nu \rangle} \in \mathbb{R}^F, \nu = 1,2$. They serve as the coefficients of moments of edge $ij$, which will be introduced in next subsection.

### 1.1.4. The Rationale of Moment Interaction

We get motivation from the moment tensor descriptor[4] for atom $i$, which is defined as a summation from edge moments as $\sum_{j \in \mathcal{N}(i)} f_{ij}^{\langle \nu \rangle}(r_{ij}) \vec{r}_{ij}^{\otimes \nu}$, which encapsulated radial part $f_{ij}^{\langle \nu \rangle}(r_{ij})$ and angular part $\vec{r}_{ij}^{\otimes \nu}$ containing angular information about the neighborhood $\mathcal{N}(i)$ and is tensor of rank $\nu$.[4]

Our MGNN defines the moment of an edge $ij$ as $f_{ij}^{\langle \nu \rangle}(r_{ij}, z_i, z_j) \hat{r}_{ij}^{\otimes \nu}$ to account for both $r_{ij}$ and the charges $z_i$ and $z_j$, thereby enriching the function's capacity to encapsulate the physicochemical context. Note that while $f_{ij}^{\langle \nu \rangle}(r_{ij})$ is originally a scalar function of $r_{ij}$ in **Ref.**[4], in MGNN $f_{ij}^{\langle \nu \rangle}(r_{ij}, z_i, z_j)$ is approximated using neural network.

As the edge moments are contracted to formulate triplet information, edge features (i.e., $f_{ij}$ in **Equation (8)**, **(9)**, **(11)**, and **(13)**) play the role of coefficients to multiply with triplet moment (i.e., $M_{\theta_{jik}}^{\langle \nu \rangle}$ in **Equation (12)**) to construct the message passed by triplet (i.e., $m_{jik}^{\langle \nu \rangle}$ in **Equation (11)**). Within a triplet (i.e., **Figure 1**B) composed of nodes $i, j, k$ and edges $ij$, $ik$, for $\nu = 1$, $\hat{r}_{ij}^{\otimes \nu} = \hat{r}_{ij}$, so that $M_{\theta_{jik}}^{\langle 1 \rangle} = \langle \hat{r}_{ij}, \hat{r}_{ik} \rangle = \cos(\angle \hat{r}_{ij}, \hat{r}_{ik})$, and for $\nu = 2$, $\hat{r}_{ij}^{\otimes \nu} = \hat{r}_{ij} \otimes \hat{r}_{ij}$, so that $M_{\theta_{jik}}^{\langle 2 \rangle} = \langle \hat{r}_{ij}^{\otimes 2}, \hat{r}_{ik}^{\otimes 2} \rangle = \cos^2(\angle \hat{r}_{ij}, \hat{r}_{ik})$ (i.e., the Frobenius product). The coefficients $f_{ij}^{\langle \nu \rangle}$ and $f_{ik}^{\langle \nu \rangle}$ for edges $ij$ and $ik$ are derived from **Equation (13)**.

*1.1.5. Output Head*

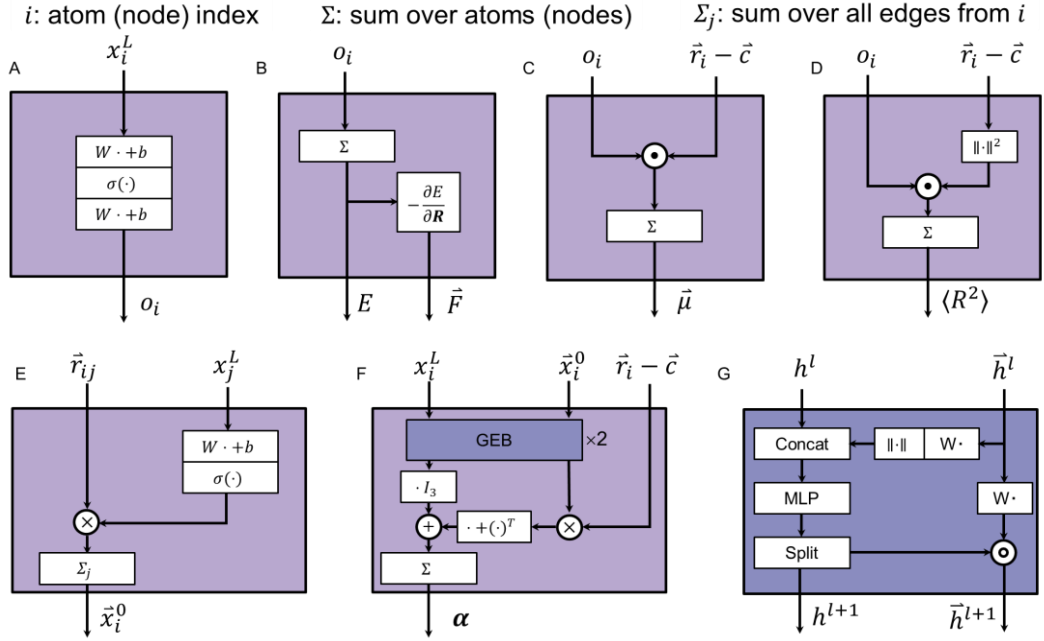

**Figure 2.** Output blocks. (A), the MLP block. (B), the energy and force. (C), dipole moment. (D), electronic spatial extent. (E), vectorial features initialization. (F), polarizability, the "GEB×2" denote there are two gated equivariant blocks (GEBs). (G), the gated equivariant block.

When a scalar is needed, the last feedforward network transforms feature on each node into a scalar (**Equation (14)**), as shown in **Figure 2**A. We perform a sum aggregation over all nodes to predict scalar quantities like energy (**Equation (15)**), as shown in **Figure 2**B.

$$o_i = MLP(x_i^L) \tag{14}$$

$$E = \sum_i o_i \tag{15}$$

The force acting on atoms $i$, $\vec{F}_i$, is computed using auto-differentiation according to its definition as the negative gradient of the total energy with respect to the position of atom $i$:

$$\vec{F}_i(Z_1, \ldots, Z_n, \vec{r}_i, \ldots, \vec{r}_n) = -\frac{\partial E}{\partial \vec{r}_i}(Z_1, \ldots, Z_n, \vec{r}_i, \ldots, \vec{r}_n) \tag{16}$$

which gives an energy-conserving force field (**Equation (16)**), as shown in **Figure 2**B.[34]

The molecular dipole moment $\vec{\mu}$ is the response of the molecular energy to an electric field $\nabla_{\vec{F}} E$ and, at the same time, the first moment of the electron density.[19] It is often predicted using latent atomic charges, and in our model, we predict the dipole moment using **Equation (17)** and the electronic spatial extent by **Equation (18)**, as shown in **Figure 2**C and D respectively, consistent with PaiNN.[19]

$$\vec{\mu} = \sum_i o_i(\vec{r}_i - \vec{c}) \tag{17}$$

$$\langle R^2 \rangle = \sum_i o_i(\|\vec{r}_i - \vec{c}\|)^2 \tag{18}$$

This assumes the center of mass at $\vec{c}$ for brevity.

Specifically, when it is necessary to predict properties related to vectors or tensors, a gated equivariant block (for more information about rotational invariance and equivariance, please refer to Supporting Information S5),[19,35] as shown in **Figure 2**F, is needed. If one needs to calculate the infrared and Raman spectra of a molecular system using deep learning potential functions, it is also necessary to compute the system's dipole moment and polarizability. In practical experiments, we utilize **Equation (17)** to predict the system's dipole moment. However, to predict the molecular polarizability tensor, we require a vector feature to construct the tensor form. Therefore, we first initialize the node's vector feature in the output head:

$$\vec{x}_i^0 = \sum_{j \in \mathcal{N}(i)} \sigma(W x_j^L) \otimes \vec{r}_{ij} \tag{19}$$

where $W \in \mathbb{R}^{F \times F}$ is the learnable weight used to transform destination node feature to control how much the $\vec{r}_{ij}$ contributes to the initial node vectorial features. $x_j^L \in \mathbb{R}^F$ is the destination node representation of the edge $ij$, and the superscript $L$ denotes it is the output of the last interaction layer. $\vec{r}_{ij} \in \mathbb{R}^3$ is the vector in Cartesian coordinates from source node to destination node. Then through **Equation (19)** vectorial features $\vec{x}_i^0 \in \mathbb{R}^{F \times 3}$ of atom $i$ can be initialized, in which the superscript $0$ denotes the initial vectorial features, as shown in **Figure 2**E.

Considering the prediction of polarizability tensor (i.e., the response of the molecular dipole to an electric field[19]), we construct polarizability tensors using

$$\boldsymbol{\alpha} = \sum_{i=1}^{N} \alpha_0(x_i^L) I_3 + \vec{v}(\vec{x}_i) \otimes (\vec{r}_i - \vec{c}) + (\vec{r}_i - \vec{c}) \otimes \vec{v}(\vec{x}_i) \tag{20}$$

where $I_3$ is the identical matrix with shape $3 \times 3$, and $\vec{v}(\vec{x}_i)$ transforms vector features $(\vec{x}_i)$ initialized by **Equation (19)** and passing them through two gated equivariant blocks[19,35], each yielding atom-wise scalars and vectors. The first term describes isotropic, atom-wise polarizabilities. The other two terms are the anisotropic components of the polarizability tensor. The atom positions subtracting the center of mass $\vec{r}_i - \vec{c}$ are used here to incorporate the global structure of the molecule.[19] The calculation of polarizability tensor $\boldsymbol{\alpha}$ is illustrated in **Figure 2**F and the gated equivariant block in **Figure 2**G.

We utilize the open-source deep learning frameworks PyTorch 2.0.1[36], SchNetPack 2.0[37], PyTorch Geometric 2.3.1[38], and PyTorch Lightning 2.1.3 for constructing and training MGNN (see more details in Supporting Information S3). Considering both moment orders $\nu = 1$ and $\nu = 2$, has better performance than considering $\nu = 1$ alone (details in Supporting Information S4).

# Benchmark

We show that our MGNN architecture can achieve new state-of-the-art results on popular benchmarks like the QM9 dataset[26,27] and the revised MD17 dataset[28]. We refer the reader to the Supporting Information S3 for further training, data set and experimental details.

## 1.2. QM9

The QM9 dataset serves as a widely recognized benchmark for predicting various molecular properties at equilibrium.[26,27] It comprises 130831 optimized structures with chemical elements (C, H, O, N, F) containing up to 9 heavy atoms (C, O, N, F). In our benchmarking of MGNN, we target 12 properties. To maintain consistency with prior research, MGNN is trained on 110000 examples, while 10000 molecules are reserved for validation purposes to regulate learning rate decay and enable early stopping. The remaining data is allocated as the test set, and results are averaged over three random splits. **Table 1** presents a comparison with several state-of-the-art methods that learn the properties described above through a direct mapping from atomic coordinates and species. SchNet[21], Physnet[24], Dimenet++[25], Equiformer[39], PaiNN[19], Allegro[12], SphereNet[40], TorchMD-Net[41], and TensorNet[42] are included in **Table 1** for reference. MGNN attains state-of-the-art performance in seven target properties and delivers comparable results in the remaining targets.

**Table 1**. MAE results on QM9 testing set for various chemical properties. Results for MGNN are averaged over three random splits. "-" denotes no results are reported in the referred papers for the corresponding properties. Best in **bold**.

| Property | Unit | SchNet | Physnet | Dimenet++ | Equiformer | PaiNN | Allegro | SphereNet | TorchMD-NET | TensorNet | MGNN |
|---|---|---|---|---|---|---|---|---|---|---|---|
| $\mu$ | $D$ | 0.033 | 0.053 | 0.03 | 0.011 | 0.012 | - | 0.026 | 0.011 | - | **0.01** |
| $\alpha$ | $a_0^3$ | 0.235 | 0.062 | 0.044 | 0.046 | 0.045 | - | 0.046 | 0.059 | - | **0.041** |
| $\epsilon_{HOMO}$ | $meV$ | 41 | 32.9 | 24.6 | **15** | 27.6 | - | 23 | 20 | - | 23.2 |
| $\epsilon_{LUMO}$ | $meV$ | 34 | 24.7 | 19.5 | **14** | 20.4 | - | 18 | 18 | - | 17 |
| $\Delta\epsilon$ | $meV$ | 63 | 42.5 | 32.6 | **30** | 45.7 | - | 32 | 36 | - | **30**[a] |
| $\langle R^2 \rangle$ | $a_0^2$ | 0.073 | 0.765 | 0.331 | 0.251 | 0.066 | - | 0.292 | **0.033** | - | 0.04 |
| ZPVE | $meV$ | 1.7 | 1.39 | 1.21 | 1.26 | 1.28 | - | **1.12** | 1.84 | - | 1.17 |
| $U_0$ | $meV$ | 14 | 8.15 | 6.32 | 6.59 | 5.85 | 4.7 | 6 | 6.15 | 4.3 | **4.1** |
| $U$ | $meV$ | 19 | 8.34 | 6.28 | 6.74 | 5.83 | 4.4 | 7 | 6.38 | 4.3 | **4.2** |
| $H$ | $meV$ | 14 | 8.42 | 6.53 | 6.63 | 5.98 | 4.4 | 6 | 6.16 | 4.3 | **4.1** |
| $G$ | $meV$ | 14 | 9.4 | 7.56 | 7.63 | 7.35 | **5.7** | 8 | 7.62 | 6 | **5.7** |
| $C_v$ | $\frac{cal}{mol\,K}$ | 0.033 | 0.028 | 0.023 | 0.023 | 0.024 | - | **0.021** | 0.026 | - | 0.023 |

[a] Note that we predict $\Delta\epsilon$ simply by computing $\epsilon_{LUMO} - \epsilon_{HOMO}$, as this is precisely how it is calculated in by DFT calculations.

## 1.3. Revised MD17

In our evaluation on the revised MD17 dataset[28], which contains 100000 recomputed structures of 10 molecules from original MD17 dataset[34,43,44], we assess MGNN's capability to accurately learn the energies and forces of small molecules. This dataset comprises ten small, organic molecules, and we aim to predict both energy and forces on systems with fixed chemical composition exhibiting conformational changes. For each molecule, we allocate 950 configurations for training and 50 for validation, uniformly sampled from the entire dataset. Test error is evaluated on all remaining configurations in the dataset. A distinct model is trained for each trajectory. **Table 2** presents a comparison with GAP[2], ACE[5], GetNet-(T/D)[22], NequIP(l=3)[11], Allegro[12], which are reported in **Ref.**[12], and TensorNet 3L[42], BOTNet[45], and MACE[46] which are reported in **Ref.**[42]. MGNN achieves state-of-the-art results for the molecules Ethanol, Benzene, and Uracil, and notably sets a new state-of-the-art for the energy MAE of Ethanol. Furthermore, in other molecular systems, MGNN surpasses previous descriptor-based machine learning interatomic potentials (GAP, ACE), invariant GNNs(GemNet-(T/Q)) or achieves comparable results to other equivariant GNNs(NequIP(l=3),

Allegro, TensorNet 3L, BOTNet, MACE).

**Table 2**. MAE results on revised MD17 dataset for energy and force predictions in units of [meV] and [meV Å$^{-1}$], respectively. Best in **bold**.

| Molecule | | GAP | ACE | GemNet-(T/Q) | NequIP | Allegro | Tensor Net | BOTNet | MACE | MGNN |
|---|---|---|---|---|---|---|---|---|---|---|
| Aspirin | E | 17.7 | 6.1 | – | 2.3 | 2.3 | 2.4 | 2.3 | **2.2** | 3.1 |
| | F | 44.9 | 17.9 | 9.5 | 8.2 | 7.3 | 8.9 | 8.5 | **6.6** | 9.1 |
| Azobenzene | E | 8.5 | 3.6 | – | **0.7** | 1.2 | **0.7** | **0.7** | 1.2 | 2.3 |
| | F | 24.5 | 10.9 | – | 2.9 | **2.6** | 3.1 | 3.3 | 3 | 6.2 |
| Benzene | E | 0.75 | 0.04 | – | 0.04 | 0.3 | **0.02** | 0.03 | 0.4 | 0.03 |
| | F | 6 | 0.5 | 0.5 | 0.3 | **0.2** | 0.3 | 0.3 | 0.3 | **0.2** |
| Ethanol | E | 3.5 | 1.2 | – | 0.4 | 0.4 | 0.5 | 0.4 | 0.4 | **0.3** |
| | F | 18.1 | 7.3 | 3.6 | 2.8 | **2.1** | 3.5 | 3.2 | **2.1** | 2.7 |
| Malonaldehyde | E | 4.8 | 1.7 | – | 0.8 | **0.6** | 0.8 | 0.8 | 0.8 | 0.8 |
| | F | 26.4 | 11.1 | 6.6 | 5.1 | **3.6** | 5.4 | 5.8 | 4.1 | 5.1 |
| Naphthalene | E | 3.8 | 0.9 | – | **0.2** | 0.5 | **0.2** | **0.2** | 0.5 | 0.5 |
| | F | 16.5 | 5.1 | 1.9 | 1.3 | **0.9** | 1.6 | 1.8 | 1.6 | 2.6 |
| Paracetamol | E | 8.5 | 4 | – | 1.4 | 1.5 | 1.3 | **0.3** | 1.3 | 2.4 |
| | F | 28.9 | 12.7 | – | 5.9 | 4.9 | 5.9 | 5.8 | **4.8** | 6.8 |
| Salicylic acid | E | 5.6 | 1.8 | – | **0.7** | 0.9 | 5.9 | 0.8 | 0.9 | 1.3 |
| | F | 24.7 | 9.3 | 5.3 | 4 | **2.9** | 4.6 | 4.3 | 3.1 | 6.4 |
| Toluene | E | 4 | 1.1 | – | **0.3** | 0.4 | **0.3** | 0.4 | 0.5 | 0.4 |
| | F | 17.8 | 6.5 | 2.2 | 1.6 | 1.8 | 1.7 | 1.9 | **1.5** | 2.4 |
| Uracil | E | 3 | 1.1 | – | **0.4** | 0.6 | **0.4** | **0.4** | 0.5 | **0.4** |
| | F | 17.6 | 6.6 | 3.8 | 3.1 | **1.8** | 3.1 | 3.2 | 2.1 | 2.7 |

# Applications

We demonstrate MGNN's advantage as molecular potential for predicting molecular dynamics simulations of amorphous electrolytes, and rapid simulation of molecular spectra in vacuum conditions for organic molecules. We refer the reader to the Supporting Information S3 for further training, data set, experimental details, and molecular dynamics simulation settings.

## 1.4. Li-ion Diffusion in A Phosphate Electrolyte

Our investigation into the MGNN's capabilities extends to the kinetic properties of Li-ion diffusion in $Li_3PO_4$ solid electrolyte, a material whose conductivity is intricately linked to its crystallinity. The $Li_3PO_4$ structure consists of 192 atoms. The simulations are conducted using a time step of 2 fs in the NVT ensemble, employing Nosé-Hoover thermostat[47]. The dataset includes a 50 ps ab-initio molecular dynamic (AIMD) simulation in the molten liquid state as T = 3000 K, followed by a 50 ps AIMD simulation in the quenched state at T = 600 K. These trajectories are combined, and both the training set of 10000 structures and the validation set of 1000 are randomly sampled from the combined dataset of 50000 structures.

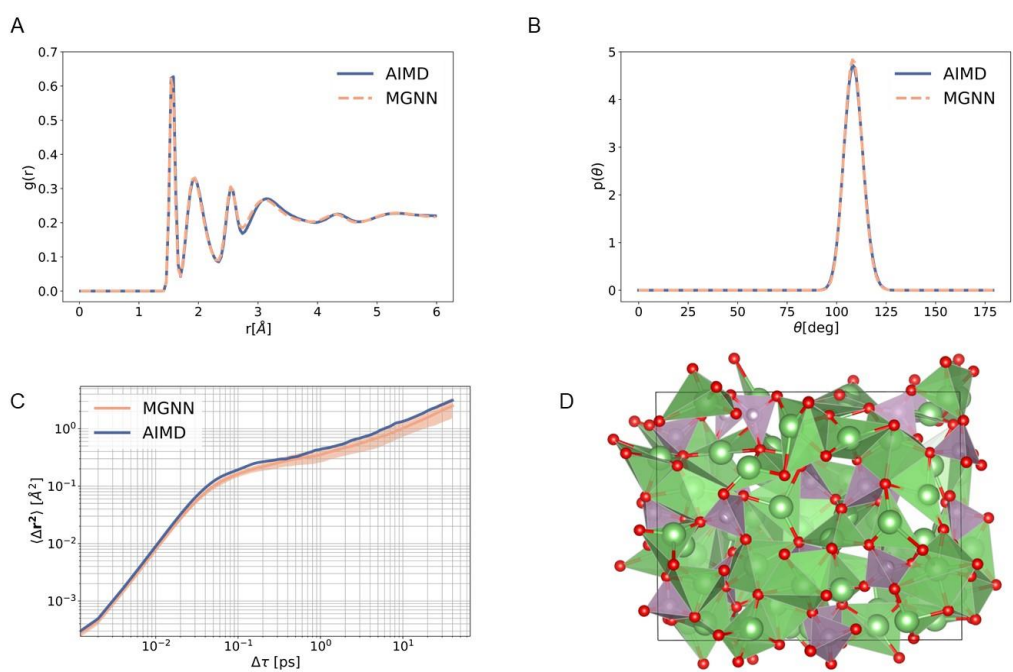

**Figure 3.** Structure properties of $Li_3PO_4$ and Li dynamic in $Li_3PO_4$. (A) Radial distribution function. (B)

Angular distribution function of tetrahedral bond angle. All defined probability density functions. (C) Comparison of the Li MSD of AIMD vs. MGNN. Results are averaged over 10 runs of MGNN, shading indicates +/- one standard deviation. (D) The quenched $Li_3PO_4$ structure at T = 600 K. The VESTA[48] software was used to generate this figure.

**Table 3**. Energy and Force MAE for the $Li_3PO_4$ and Ag datasets, reported in units of [meV atom$^{-1}$] and [meV Å$^{-1}$]. "E" denotes energy and "F" forces.

| Dataset | | Allegro[12] | MGNN |
|---|---|---|---|
| $Li_3PO_4$ | E | 1.7 | **0.2** |
| | F | 73.4 | **22.9** |
| Ag[a)] | E | 0.397 | **0.168** |
| | F | 16.8 | **9.5** |

[a)] To maintain consistency, we not only present the data for $Li_3PO_4$ in the table but also include another dataset demonstrated in their paper[12], namely, the Ag bulk crystal with a vacancy, simulated at 90% of the melting temperature. For further details, please refer to **Ref.**[12] and our Supporting Information S3.

Note that the reference results for Allegro in **Table 3** are sourced from their paper[12]. The Ag model is trained and validated using 1000 structures, consist with **Ref.**[12], resulting in a final training performance with energy mean absolute error (MAE) of 0.168 meV atom$^{-1}$ and force component MAE of 9.5 meV Å$^{-1}$ on a test set comprising 159 structures. For both the $Li_3PO_4$ and Ag datasets, we have achieved competitive results comparable to Allegro.

When evaluating the $Li_3PO_4$ test set for the quenched amorphous state, which the simulation is conducted on, we obtained a MAE in the energies of 0.2 meV atom$^{-1}$, along with a MAE in the force components of 22.92 meV Å$^{-1}$. Subsequently, we conduct a series of ten MD simulations starting from the initial structure of the quenched AIMD simulation, each lasting 50 ps at T = 600 K in the quenched state, to scrutinize how well MGNN captures the structure and kinetics compares to AIMD. To assess the quality of the structure after the phase change, we compare the all-atom radial distribution functions (RDF) and the angular distribution functions (ADF) of the tetrahedral angle O-P-O (P central atom). As depicted in **Figure 3**A, B, MGNN adeptly reproduces both distribution functions. Concerning ion transport kinetics, we evaluate how accurately MGNN models the Li mean-square-displacement (MSD) in the quenched state. Once more, we observe excellent agreement with AIMD, as shown in **Figure 3**C. The structure of $Li_3PO_4$ is depicted in **Figure 3**D. Reference data for the $Li_3PO_4$ is obtained from **Ref.**[12].

Molecular dynamics simulations are executed in LAMMPS[49] using the MGNN pair style, implemented in the SchNetPack 2.0[37] interface. Other details refer to Supporting Information S3.4. The RDF and ADF for $Li_3PO_4$ are computed with a maximum distance of 6 Å (RDF) and 2.5 Å (ADF). The simulation commences from the initial frame of the AIMD quenched simulation. RDF and ADF for MGNN are averaged over ten runs with different initial velocities, with the first 10 ps of the 50 ps simulation discarded in the RDF/ADF analysis to ensure equilibration is adequately accounted for.

## 1.5. Molecular Spectra

MGNN is also employed to efficiently compute the infrared and Raman spectra of ethanol. Traditional methods, which might rely on the harmonic oscillator approximation from a singular molecular structure, often neglect critical phenomena such as the diversity of molecular conformations. To secure spectra of superior quality, it is imperative to undertake molecular dynamics simulations, which demand the prediction of forces at each timestep and the calculation of dipole moments and polarizabilities throughout the trajectory. The acquisition of infrared and Raman spectra through the Fourier transformation of time autocorrelation functions of dipole moments and polarizabilities respectively alleviates the complexity and computational intensity of using electronic structure methods. Given the necessity to account for nuclear quantum effects (NQE) for high-fidelity spectra, ring-polymer molecular dynamics (RPMD) simulations are employed, managing multiple replicas of the molecule simultaneously, thereby significantly increasing the computational burden.[19]

A joint model has been trained on a dataset comprising 8,000 conformations of ethanol, supplemented by an additional 1,000 molecules each for validation and testing purposes. This model is tasked with the prediction of energies, forces, dipole moments, and polarizability tensors. Dipole moments and polarizability tensors are determined as described in **Equation (17)** and **(20)**, respectively. The unified model demonstrates precise predictions for energy, forces, dipole moments, and polarizabilities, as demonstrated in **Table 4**. Reference data for the ethanol molecule is sourced from **Ref.**[20], and the experimental spectra recorded in the gas phase are obtained from **Ref.**[50,51].

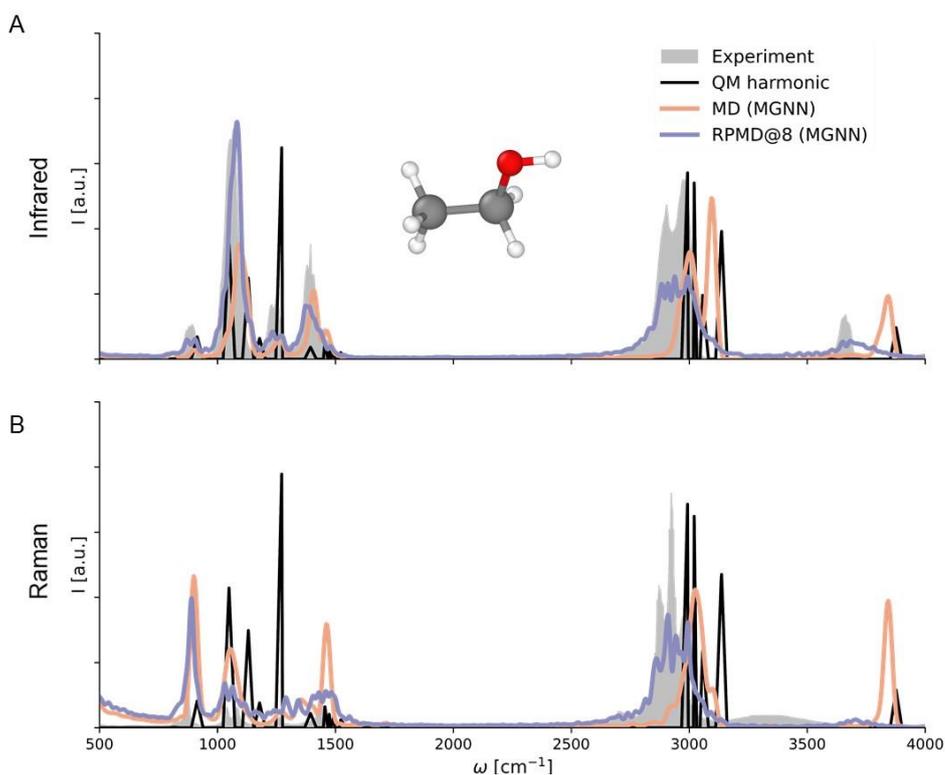

**Fig. 4.** Infrared (A) and Raman (B) spectra of ethanol. Spectra calculated with the reference method using the harmonic oscillator approximations are shown in black (QM harmonic). Experimental results are shown in gray.

**Table 4**: Ethanol in vacuum results. Mean absolute error for the prediction of energies (E), forces (F), dipole moments ($\vec{\mu}$), and polarizabilities ($\alpha$), averaged over random 3 splits, with corresponding units between parentheses.

| Model | E [kcal mol$^{-1}$] | F [kcal mol$^{-1}$ Å$^{-1}$] | $\vec{\mu}$ [D] | $\alpha$ [Bohr$^3$] |
|---|---|---|---|---|
| PaiNN[19] | 0.027 | 0.150 | 0.003 | 0.009 |
| FieldSchNet[20] | 0.017 | 0.128 | 0.004 | 0.008 |
| TensorNet[42] | 0.008 | 0.058 | **0.003** | **0.007** |
| MGNN | **0.006** | **0.012** | 0.003 | 0.008 |

Results from **Table 4** reveal that MGNN can acquire expressive atomic tensor embeddings, enabling simultaneous prediction of multiple molecular properties. Particularly noteworthy is MGNN's superior performance, with energy and force errors approximately a factor of two and three smaller, respectively, compared to FieldSchNet[20] and PaiNN[19]. Additionally, MGNN achieves smaller MAE in force component prediction compared to the equivariant neural network TensorNet[42], while attaining results comparable to TensorNet for dipole moments and polarizabilities.

The MGNN potential is then used for classical MD and RPMD[52]. While both methods can calculate the spectra, they focus on different aspects. The peak positions and intensities of the ethanol infrared spectra (**Figure 4**A) computed using classical MD and MGNN closely align with the static electronic structure reference, particularly evident in the C-H and O-H stretching regions at 3000 cm$^{-1}$ and 3900 cm$^{-1}$. This indicates MGNN's faithful reproduction of the original electronic structure method. However, in comparison to experimental data, both spectra exhibit shifts towards higher frequencies. In contrast, the MGNN RPMD spectrum demonstrates excellent agreement with experimental results, highlighting the significance of NQEs. Similar trends are observed for the ethanol Raman spectrum (**Figure 4**B), where the MGNN RPMD spectrum once again provides the most faithful reproduction of experimental observations.

More simulation details are shown in Supporting Information S3.6.

## Conclusion

In this work, we have introduced MGNN, a Graph Neural Network framework specifically designed for the rapid and accurate prediction of molecular properties. Our manuscript not only showcases MGNN's new state-of-the-art performance on benchmark datasets such as QM9 and revised MD17 but also its remarkable consistency with ab-initio calculations, particularly in modeling the migration energy of lithium ions in electrolytes and the calculations of infrared and Raman spectra of ethanol. MGNN's output module is adept at interfacing with both vector and tensor modules, which enables the model to tailor its output representation to the property being predicted. While our approach does not incorporate vector computations within the molecular representation learning process, the inclusion of these computations in the final output module significantly enhances the model's computational efficiency. This is exemplified in our application of MGNN to predict vector and tensor features of molecules, such as the dipole moment and polarizability, which are crucial for computing the infrared and Raman spectra of molecules like ethanol in vacuum conditions.

Looking ahead, the potential applications of MGNN are vast and varied, extending to areas such as molecular generation and wavefunction prediction. The rapid and accurate prediction of molecular properties is contingent upon highly precise reference methods, which often come with a high computational cost. As such, machine learning models like MGNN stand to play a pivotal role in surmounting these challenges, thereby driving forward the frontier of computational chemistry and materials science research.